# Single-Carrier Transport in Graphene/hBN Superlattices

*Takuya Iwasaki,\*,1 Shu Nakaharai,2 Yutaka Wakayama,2 Kenji Watanabe,3 Takashi Taniguchi,3 Yoshifumi Morita,4 and Satoshi Moriyama†,2*

[1]International Center for Young Scientists (ICYS), National Institute for Materials Science (NIMS), Tsukuba, Ibaraki 305-0044, Japan. [2]International Center for Materials Nanoarchitectonics (WPI-MANA), NIMS, Tsukuba, Ibaraki 305-0044, Japan. [3]Research Center for Functional Materials, NIMS, Tsukuba, Ibaraki 305-0044, Japan. [4]Faculty of Engineering, Gunma University, Kiryu, Gunma 376-8515, Japan.

Email: \*IWASAKI.Takuya@nims.go.jp, †MORIYAMA.Satoshi@nims.go.jp

ABSTRACT: Graphene/hexagonal boron nitride (hBN) moiré superlattices have attracted interest for use in the study of many-body effects and fractal physics in Dirac fermion systems. Many exotic transport properties have been intensively examined in such superlattices, but previous studies have not focused on single-carrier transport. The investigation of the single-carrier behavior in these superlattices would lead to an understanding of the transition of single-particle/correlated phenomena. Here, we show the single-carrier transport in a high-quality bilayer graphene/hBN superlattice-based quantum dot device. We demonstrate remarkable device controllability in the energy range near the charge neutrality point (CNP) and the hole-side satellite point. Under a perpendicular magnetic field, Coulomb oscillations disappear near the CNP, which could be a signature of the crossover between Coulomb blockade and quantum Hall regimes. Our results pave the way for exploring the relationship of single-electron transport and fractal quantum Hall effects with correlated phenomena in two-dimensional quantum materials.

Graphene stacked on hexagonal boron nitride (hBN) with the crystallographic axis alignment generates a long-period superlattice potential because of a lattice-constant mismatch of ~1.8%.[1-7] This moiré superlattice breaks inversion symmetry and modifies the electronic band structure of graphene, leading particularly to an energy gap at the charge neutrality point (CNP) and the emergence of satellites of the CNP. Previous studies have intensively examined the exotic transport properties in graphene/hBN superlattice systems, such as fractal quantum Hall effects,[1-3] topological valley Hall effects,[8-10] and correlated superconductivity.[11,12] On the other hand, single-carrier transport properties in graphene have been investigated by employing graphene-based quantum dot (QD) devices.[13-25]

To form a potential barrier for graphene QDs, geometric patterning has been performed on nanostructures by etching[13-19] or applying a perpendicular electric field to bilayer graphene (BLG).[20-25] For etching, the graphene channel must be narrowed to a few tens of nanometers to obtain a sizable gap.[13-18,26] Consequently, the carrier transport property is dominated by the unintentional carrier localization induced by edge disorders, leading to an unstable dot configuration.[13-16] Additionally, the previous devices fabricated using graphene flakes on $SiO_2$ substrates suffered from strong charge inhomogeneity due to surface roughness and substrate impurities,[13–18] which further disturbed the dot definition. On the other hand, with recent progress in fabrication technologies such as hBN encapsulation and the fabrication of a graphite back gate[3] through an all-dry transfer process,[27] high-quality BLG-based QD devices defined by electrostatically-induced potential barrier have been achieved under a perpendicular electric field.[22-25]

In this work, to study the single-carrier transport in graphene superlattices, we fabricated a BLG/hBN moiré superlattice-based QD device (Figure 1c-e), in which the double dots are defined by geometric patterning and local gating with the contribution of the superlattice potential. Note that widths of constrictions and diameters of the dots were slightly large compared to the previous graphene QD devices,[13-18] for avoiding strong potential fluctuation which leads to unstable dot configuration. We demonstrate highly controllable single/double dot operation in the range not only near the CNP, but also near the hole-side satellite. Furthermore, near the CNP, the Coulomb blockade effect is found to be suppressed under a perpendicular magnetic field, which could be a signature of the collapse of single-electron tunneling in the quantum Hall regime.

To elucidate the influence of the superlattice potential, we fabricated a Hall bar close to the double dots on the same heterostructure, as shown in Figure 1a,b, and measured them in the same environment. Note that the devices were fabricated in a region where interfacial bubbles were absent, which was achieved through a bubble-free transfer technique.[28] The devices consist of one-dimensional edge-contacting electrodes[27] and the graphite back-gate. The device fabrication process is detailed in Supporting Information. All transport measurements were performed in a dilution refrigerator with the bath temperature of 40 mK. The effective electron temperature was ~160 mK estimated by fitting of the Coulomb peak with the orthodox Coulomb blockade model[29] (shown in Supporting Information). The double-dot device was measured in two-terminal DC configuration, while the Hall bar device was measured using AC lock-in techniques with an excitation current of 10 nA and a frequency of 17 Hz. A superconducting magnet was used to apply a perpendicular magnetic field up to 6 T.

To confirm the superlattice property in the double dots, we first study the transport characteristics of the Hall bar. Figure 1f shows a map of the longitudinal resistivity $\rho$ of the Hall bar as a function of the back-gate voltage $V_{BG}$ and a perpendicular magnetic field $B$. We observe well-resolved constant-filling-factor regions corresponding to the filling factor $v = \pm 4N_{LL}$ ($N_{LL}$ is the Landau level (LL) index). These regions fan out from $V_{BG} \sim 0$ V where the carrier type is switched, indicating that the CNP is located at $V_{BG} \sim 0$ V. In addition, lines fanning out from $V_{BG} \sim \pm 15$ V are observed, which correspond to the second-generation Landau fan, and its origin at $B = 0$ T can be identified as the satellites.[1–3] From the distance between the CNP and the satellites, we estimate the moiré wavelength to be $\lambda \sim 10.7$ nm,[2] which corresponds to a BLG/hBN crystallographic-axis alignment angle of $\theta \sim 0.85°$.[4] Moreover, as shown in Figure 1g, the Hall conductivity $\sigma_{xy}$ at $B = 6$ T exhibits plateaus for $\sigma_{xy} = 4N_{LL}(e^2/h)$, where $e$ is the electron charge and $h$ is Planck's constant. These results confirm that the base heterostructure comprises a high-quality BLG/hBN superlattice[2] (detailed device characterization of the Hall bar is presented in Supporting Information).

Figure 2a shows the drain current $I_D$ as a function of $V_{BG}$ with a fixed source-drain bias voltage $V_{SD} = 100$ μV for the double-dot device. The suppression of background current and sharp peaks are observed in the range of $-4$ V $< V_{BG} < -1.5$ V, which is around the CNP in the Hall bar (indicated as A in Figure 2a,b). Importantly, unlike conventional graphene QD devices,[13–25] we find a dip structure at $V_{BG} \sim -13$ V in the hole transport regime, which is near the hole-side satellite in the Hall bar (indicated as B in Figure 2a,b), while no dip is found on the electron side. Periodic oscillation with a large background current is observed around this dip, as shown in Figure 2c. This is reminiscent of the interference effect of an open dot system,[30,31] which will be discussed with Figure 4. In the suppression region near the CNP, we observe double pairing peaks, as shown in Figure 2d, suggesting a Coulomb oscillation of the double dots.[17,32] Since top-gate voltage is fixed to zero for Figure 2a,c,d, this suppression region and the dip structure can be attributed mainly to the Coulomb blockade carrier transport through the double dots defined by the device geometry with the contribution from the superlattice potential. By fixing the energy at the blockade region to $V_{BG} = -3.17$ V (indicated by the arrow in Figure 2d), we draw a charge-stability diagram with the sweep of two top-gate voltages $V_{LDG}$ and $V_{RG}$ (the position of top-gates are shown in Figure 1d), as shown in Figure 2e. In this plot, three types of coupling states are observed. The first state corresponds to the horizontal lines in the region of 0.5 V < $V_{LDG}$ < 2 V and 1 V < $V_{RG}$ < 2 V, which indicate the formation of single dots. The second corresponds to the honeycomb structures in the region of −2 V < $V_{LDG}$ < −1 V and 1 V < $V_{RG}$ < 2 V, which represent strongly coupled double dots; the number of electrons in the double dots is constant inside a honeycomb. The third corresponds to apexes of the honeycomb structure, called triple points, in the region of 0.5 V < $V_{LDG}$ < 1 V and −1 V < $V_{RG}$ < 1 V (indicated by the dotted box in Figure 2e), which indicate a weakly coupled double-dot regime.[32]

In the weakly coupled double-dot regime with $V_{SD} = 200$ μV, the triple points evolve to bias triangles, as shown in Figure 3a. By applying a reverse bias of $V_{SD} = -400$ μV, the triangles are inverted, as shown in Figure 3b. From these diagrams, we extract the dot parameters by using a simple capacitance model[32] with the following parameters (see Supporting Information for details): the gate capacitance between the gate LDG (RG) and left (right) dot $C_{LDG(RG)} = e/\Delta V_{LDG(RG)} \approx 1.3$ aF (0.36 aF), conversion factor for the gate LDG (RG) to the left (right) dot $\alpha_{LDG,LD(RG,RD)} = V_{SD}/\delta V_{LDG(RG)} \approx 0.012$ (0.0041), total capacitance of the left (right) dot $C_{\Sigma,LDG(RG)} = C_{LDG(RG)}/\alpha_{LDG,LD(RG,RD)} \approx 110$ aF (90 aF), charging energy of the left (right) dot $E_C^{L(R)} \approx e/C_{\Sigma,LDG(RG)} \approx 1.5$ meV (1.8 meV), and electrostatic interdot coupling energy $E_C^m \approx \alpha_{LDG,LD}\delta V_{LDG} \approx \alpha_{RG,RD}\delta V_{RG} \approx 0.45\text{-}0.64$ meV. The identical charging energies for left and right dots could reflect the symmetric double-dot device geometry (Figure 1c). The electrostatic coupling between gate LDG and left dot is larger than that between gate RG and right dot, which is in good agreement with the device structure. By using the gate CG located between the dots (the position of the gate CG is shown in Figure 1d), we tune $E_\mathrm{C}^\mathrm{m}$ and the dot configuration to the strongly coupled double-dot regime (Figure 3c) and the single-dot-like regime (Figure 3d). Furthermore, each dot can be individually operated by tuning the gate voltage $V_\mathrm{LDG}$ and $V_\mathrm{RG}$. Figure 3e,f show Coulomb diamond plots for each dot with another top-gate voltage fixed for making another dot transparent. From these plots, the addition energies of left and right dots are estimated to be ~1.6 meV and ~1.4 meV, respectively, which are well consistent with the charging energies estimated independently from the charge-stability diagram above. The slight mismatch of the energies of the right dot could be due to the influence from the small cross-coupling. In consequence, the superlattice device shows well-defined closed double dots with remarkable controllability. Note that excited states are not observed, possibly because the single-particle level spacing is too small because of the large dot diameter of ~135 nm.[13,18]

Next, we discuss the transport property at the dip near the hole-side satellite. Figure 4a shows the charge-stability diagram drawn by fixing $V_\mathrm{BG}$ at −13.13 V. A honeycomb structure is observed, similar to the suppression region near the CNP, but with a large background current. As in the above discussion, the dot configuration can be switched between single and double dots. By tuning the gate voltages $V_\mathrm{LDG}$ and $V_\mathrm{RG}$, the oscillation of conductance ($G = I_\mathrm{D}/V_\mathrm{SD}$) can be made more periodic (upper panel in Figure 4b). Under the same condition, non-lifted Coulomb diamonds are observed (lower panel in Figure 4b). These results indicate the interference effect in the open dot system (a Fabry-Pérot interferometer), in which the dot (cavity) is strongly coupled with source and drain leads.[30,31,33] From the upper panel in Figure 4b, the average peak spacing is estimated to be $\delta V_\mathrm{BG}^\mathrm{peak} \sim 11.3$ mV. In the Fabry-Pérot cavity, resonant peaks occur under the condition $k_\mathrm{F} L_\mathrm{C} = q\pi$, where $k_\mathrm{F}$ is the Fermi wave vector, $L_\mathrm{C}$ is a cavity length, and $q$ is an integer.[31] We extract the cavity length as $L_\mathrm{C} = \pi/k_\mathrm{F} = \sqrt{\pi e/(C_\mathrm{BG}\delta V_\mathrm{BG}^\mathrm{peak})} \approx 310$ nm ($C_\mathrm{BG} \sim 43.2$ nF/cm$^2$ is the geometric back-gate capacitance estimated using a simple parallel-plate model), which is in good agreement with the distance between the top gates of ~300 nm. Note that the above open-dot properties are not clearly observed in the range away from the dip near the hole-side satellite. The absence of a dip near the electron-side satellite could reflect the high density of states in the electron side. In the case of monolayer graphene, the electron-side satellite has the threefold degeneracy of Dirac cone replicas, while the hole-side satellite has one-hold degeneracy, leading to the difference in the density of states.[7] Accordingly, we interpret that the hole-side dip induced by the superlattice potential contributes to the definition of the tunable open dot system.

Finally, we investigate the transport property of the double-dot device under a magnetic field $B$. Figure 5a shows a Landau fan diagram of the double-dot device with a large bias voltage of $V_{SD} = 10$ mV. By using the results for the Hall bar shown in Figure 1f, we extrapolate the spectral lines, which coincide with the fanning lines of the double-dot device (dotted lines in Figure 5a). The constant-filling-factor regions fanning out from $V_{BG} \sim 0$ V and from $V_{BG} \sim \pm 15$ V indicate that the positions of the CNP and the satellites are almost same for the Hall bar and the double-dot devices. Thus, the hole-side dip could be associated with the hole-side satellite stemming from the superlattice potential, supporting the above discussion. Consequently, Hofstadter's butterfly spectrum is observed in the double-dot transport with the large bias voltage that sufficiently high energy compared to the potential barriers. By applying a small bias voltage of $V_{SD} = 10$ μV, the Coulomb oscillations and current-suppression region around $V_{BG} \sim 0$ V are observed, as shown in Figure 5b. Intriguingly, we find that the suppression region expands with increasing $B$ along the lines corresponding to $v = \pm 1$ depicted by following $B = (hC_{BG}/(ve^2))V_{BG}$, and the Coulomb oscillations nearly vanish in this region ($I_D$-$V_{BG}$ curves at fixed $B$ are shown in Supporting Information). This effect is more pronounced in the hole side, as shown in Figure 5c. Because the envelope of the suppressed region follows the lines corresponding to $v = \pm 1$, this region could involve the LL gap at energy $E = 0$ ($v = 0$).[34-36] The disappearance of the Coulomb oscillations in the LL gap could manifest itself as a crossover of the Coulomb blockade and the quantum Hall regime. As the spin, valley, and orbital are degenerated in the 0th LL in BLG, the Landau gap at $E = 0$ and the lines corresponding to $v = \pm 1$ imply the complete lifting of this degeneracy, which could be related to the electron-electron interaction.[34-36] Even though the mechanism might differ from that of the BLG superlattice, we note that the gap between the $v = \pm 1$ lines may be caused by quantum Hall ferromagnetism in the case of a monolayer graphene superlattice.[7] Whereas the reason for this lifting remains an open question, we speculate that the observation arises from the high quality of BLG and the superlattice potential in our device and the resulting enhancement of the many-body effect in the sufficient low-disorder system[34-36] (the device quality in the Hall bar is shown in Supporting Information). It should be noted that, for the transport property of the double-dot device in a parallel magnetic field, the current suppression region near the CNP is not observed (shown in Supporting Information), in contrast to that in a perpendicular magnetic field (Figure 5b,c). This further confirms that the suppression region could be due to the LL gap owing to the superlattice structure as discussed above. In the vicinity of the $v = -1$ line, a few Coulomb peaks remain in the suppressed region, and they seem to be bundled into one peak as increasing $B$ (Figure 5b,c). Moreover, in the range around $V_{BG} = -1$ V away from the suppressed region, the Coulomb oscillations shift toward the negative $V_{BG}$ side for $B \leq 3$ T, and they generally shift toward the positive side for $B = 3\text{-}6$ T, as shown in Figure 5d. These behaviors resemble the Fock-Darwin state[14,15,29,37] or interdot repulsion in a QD system.[32] A detailed analysis is beyond the scope of the present paper and is a topic for future research.

In conclusion, we investigated the single-carrier transport in the BLG/hBN superlattice by realizing a highly controllable QD device. Single/double closed and open dot operations were demonstrated in the range near the CNP and the hole-side satellite, respectively. In particular, the dip induced by the superlattice Under a perpendicular magnetic field, we observed the suppression of Coulomb oscillations and the shifting/bundling of Coulomb peaks, which could be related to fractal quantum Hall and/or Fock-Darwin spectra in the massive Dirac fermion system. The Hofstadter's butterfly spectra were observed in both the Hall bar and the double-dot devices in the same heterostructure, suggesting that the observed results in the double-dot device could be due to the confined single carrier in the moiré superlattice system. For further characterization of these features in detail, the theoretical modeling/calculation and the investigation of shell-filling states are remained as future tasks. Although our device showed the high-quality and remarkable controllability, a contribution from the edge disorder due to plasma etching cannot be excluded completely, remained to be improved in future, i.e., by using electrostatic confinement induced by a perpendicular electric displacement field.[20-25] Our results show that superlattice single-electron devices could be a significant basis for investigating the interaction between exotic many-body phenomena and single-electron transport.

**Notes**

The authors declare no competing financial interest.

ACKNOWLEDGMENT

Authors thank to Hirotaka Osato, Eiichiro Watanabe and Daiju Tsuya in NIMS Nanofabrication platform for helping the device fabrication. This work was supported by Japan Society for the Promotion of Science (JSPS) KAKENHI Grant Number 19K15385, and by NIMS Nanofabrication Platform in Nanotechnology Platform Project, the World Premier International Research Center Initiative on Materials Nanoarchitectonics, sponsored by the Ministry of Education, Culture, Sports, Science and Technology (MEXT), Japan. S.M. acknowledges financial support from a Murata Science Foundation.

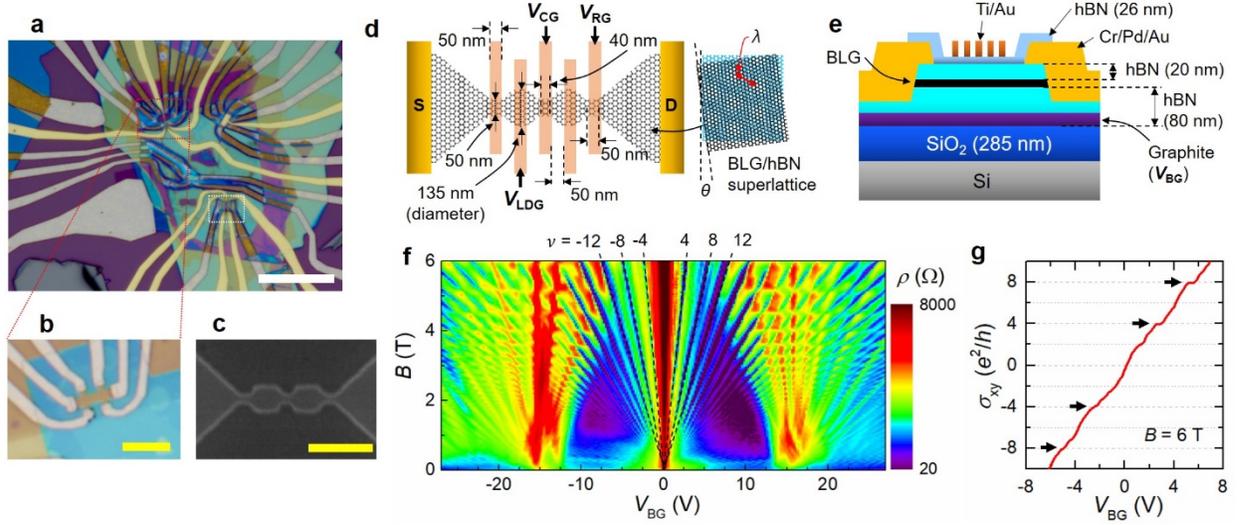

**Figure 1.** Device structure of bilayer graphene superlattice. (a,b) Optical images of the fabricated device. (a) Overall heterostructure. Scale bar: 20 μm. The dotted red and white boxes correspond to the Hall bar and the double-dot device, respectively. (b) Magnified view of the Hall bar device before top-gate fabrication. Scale bar: 5 μm. (c) Scanning electron micrograph of the identical double-dot device before top-gate fabrication. Scale bar: 300 nm. (d) Schematic top view of the double-dot device with its dimensions. S and D denote the source and drain, respectively. Right schematic image shows the bilayer graphene/hBN superlattice with a moiré wavelength $\lambda$ and a crystallographic axis angle $\theta$. (e) Schematic cross section of the double-dot device. (f) Longitudinal resistivity $\rho$ mapping of the Hall bar device in a logarithmic scale as a function of $V_{BG}$ and $B$ at $T = 40$ mK. The dashed lines depict constant-filling-factor regions for $v = \pm 4$, 8, and 12 fanning from $V_{BG} = 0$ V. (g) Hall conductivity $\sigma_{xy}$ as a function of $V_{BG}$ at $B = 6$ T. $\sigma_{xy}$ is calculated from $\sigma_{xy} = -R_{xy}/(\rho^2 + R_{xy}^2)$, where $R_{xy}$ is the transverse resistance of the Hall bar. The arrows point to plateaus for $\sigma_{xy} = 4N_{LL}(e^2/h)$.

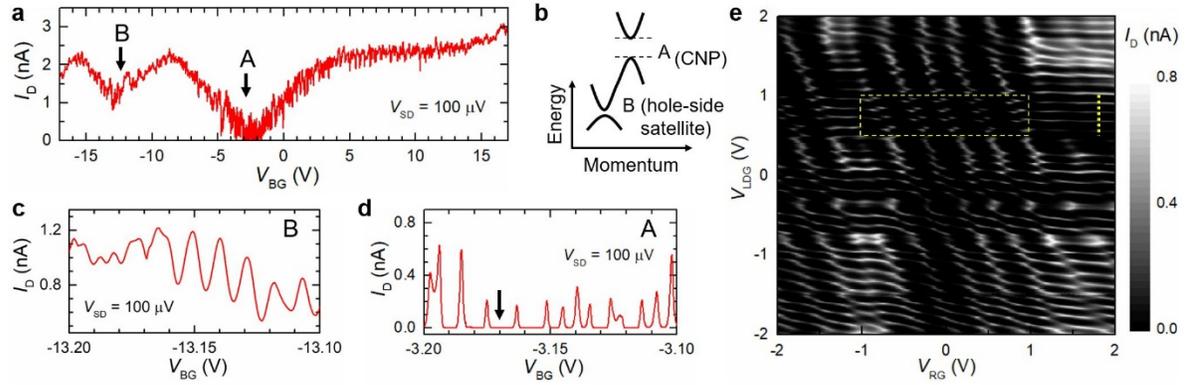

**Figure 2.** Transport properties in the double-dot device. (a) $I_D$ vs. $V_{BG}$ for $V_{SD} = 100\ \mu V$. (b) Schematic band structure of the bilayer graphene moiré superlattice. A and B indicate regions near the charge neutrality point and the hole-side satellite, respectively. (c) Magnified view of the dip region near the hole-side satellite. (d) Magnified view of the region near the charge neutrality point. (e) Charge-stability diagram ($I_D$ mapping as a function of $V_{LDG}$ and $V_{RG}$) for $V_{BG} = -3.17$ V and $V_{SD} = 100\ \mu V$. The dotted yellow box corresponds to the weakly coupled double-dot regime. The dotted yellow line indicates the range for the Coulomb diamond plot (Figure 3e).

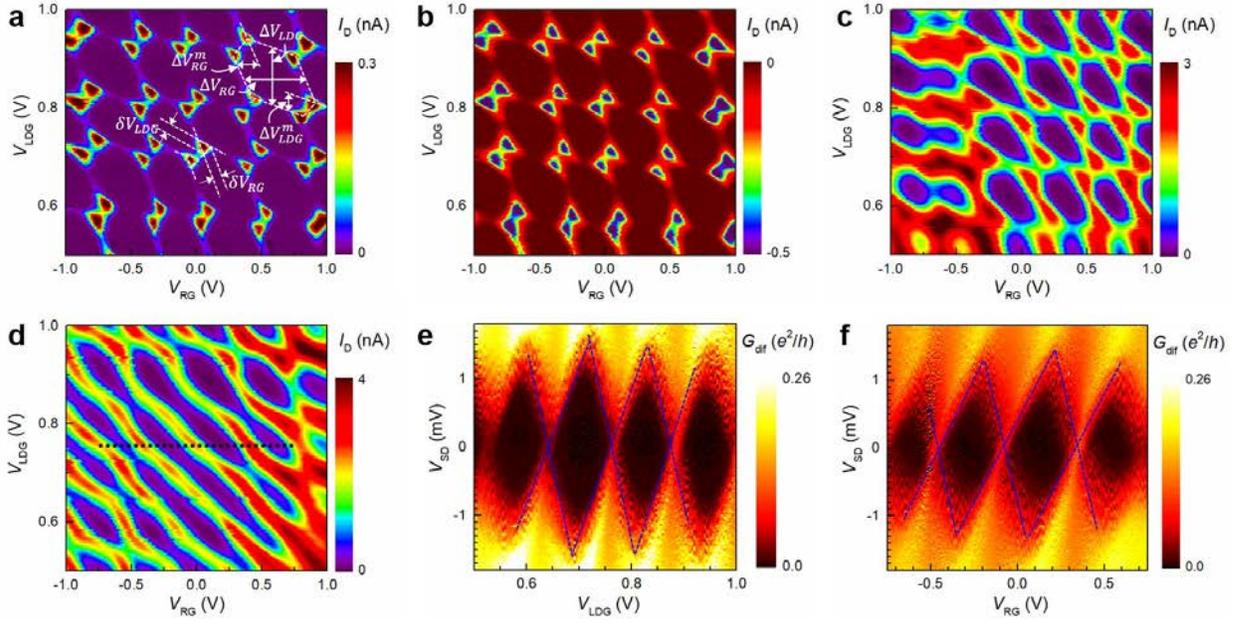

**Figure 3.** Closed dot system near the charge neutrality point. (a-d) Charge-stability diagrams with $V_{BG}$ fixed at −3.17 V. For the bias triangle plot, $V_{SD}$ = 200 μV (a) and −400 μV (b). The scales shown in a. are used for extracting the dot parameters. For tuning the interdot coupling, $V_{SD}$ = 1 mV and $V_{CG}$ = 0.154 V (c) and −0.15 V (d) are used. (e,f) Plots of differential conductance ($G_{dif}$ = $dI_D/dV_{SD}$) as a function of $V_{SD}$ and $V_{LDG(RG)}$ (Coulomb diamond plots). Individual Coulomb diamond characteristics for each dot are shown. For left-dot operation, $V_{RG}$ is fixed at 1.8 V (e). For right-dot operation, $V_{LDG}$ is fixed at 0.75 V (f). The dotted lines correspond to the boundaries of blockade regions. The measured $V_{RG}$ range is indicated by the dotted line in (d).

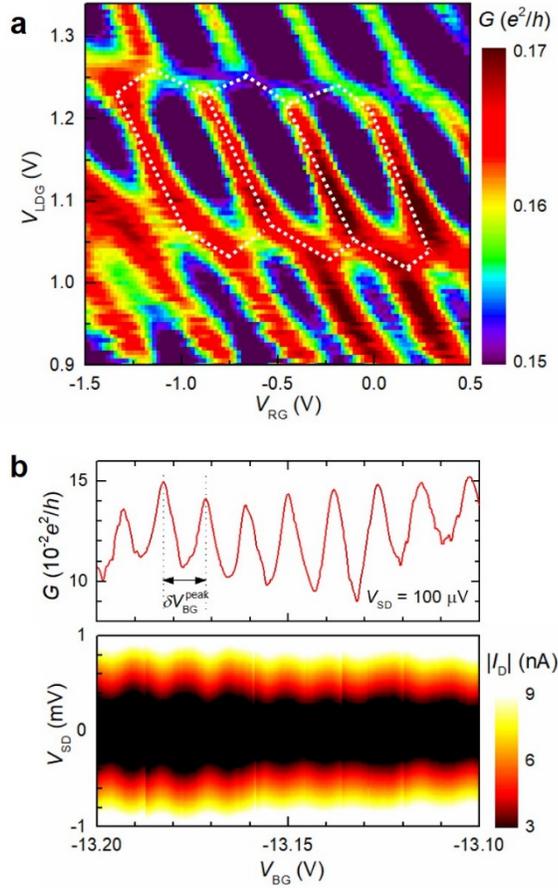

**Figure 4.** Open dot system near the hole-side satellite. (a) Charge-stability diagram (mapping of $G$ as a function of $V_{LDG}$ and $V_{RG}$) with $V_{BG}$ fixed at −13.13 V. The dotted white lines are guides to the eye for the honeycomb structures. (b) Single-dot configuration with $V_{LDG} = V_{RG} = -1.52$ V. Upper: $G$ vs. $V_{BG}$ oscillation characteristic for $V_{SD} = 100$ μV. Bottom: mapping of absolute drain current $|I_D|$ as a function of $V_{SD}$ and $V_{BG}$.

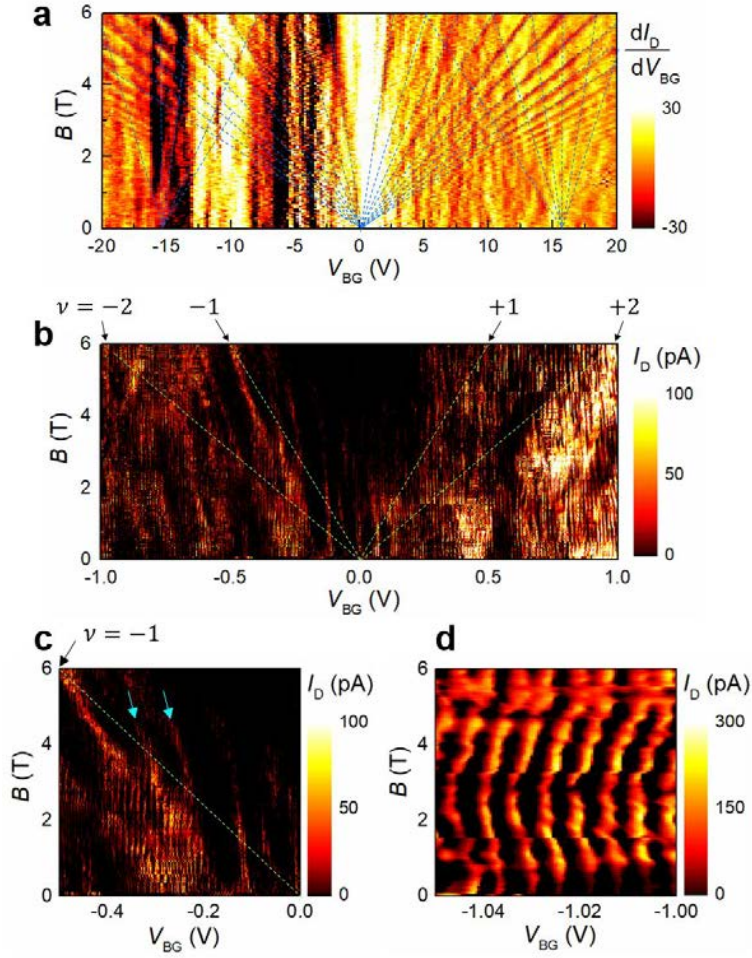

**Figure 5.** Single-carrier transport in the perpendicular magnetic field. (a) Landau fan diagram of the double-dot device for $V_{SD} = 10$ mV. $dI_D/dV_{BG}$ is used for the mapping to emphasize the constant-filling-factor regions shown as dotted blue lines. (b,c) High-resolution $I_D$ fan diagram mapping for $V_{SD} = 10$ μV near the charge neutrality point (b) and magnified view of the hole side (c). The dotted green lines are the filling-factor lines fanning from $V_{BG} = 0$ V. The light blue arrows in (c) indicate the Coulomb peak bundled with increasing $B$. (d) Evolution of the Coulomb oscillations around $V_{BG} = -1$ V with increasing $B$.

Single-Carrier Transport in Graphene/hBN Superlattices:

Supporting Information

Takuya Iwasaki,*,[1] Shu Nakaharai,[2] Yutaka Wakayama,[2] Kenji Watanabe,[3] Takashi Taniguchi,[3] Yoshifumi Morita,[4] Satoshi Moriyama[†,2]

[1] International Center for Young Scientists (ICYS), National Institute for Materials Science (NIMS), Tsukuba 305-0044, Japan.
[2] International Center for Materials Nanoarchitectonics (WPI-MANA), NIMS, Tsukuba, Ibaraki 305-0044, Japan.
[3] Research Center for Functional Materials, NIMS, Tsukuba, Ibaraki 305-0044, Japan.
[4] Faculty of Engineering, Gunma University, Kiryu, Gunma 376-8515, Japan

Email: *IWASAKI.Takuya@nims.go.jp, †MORIYAMA.Satoshi@nims.go.jp

1. Device fabrication process
2. Device characterization of the Hall bar
3. Temperature dependence of the Hall bar
4. Estimation of the effective electron temperature
5. Estimation of the double-dot parameters
6. $I_D$-$V_{BG}$ curves at the fixed perpendicular magnetic field near the charge neutrality point
7. Transport property in a parallel magnetic field

1. Device fabrication process

Bilayer graphene (BLG), graphite and hexagonal boron nitride (hBN) flakes were obtained on $SiO_2$/Si substrates by the mechanical exfoliation method using adhesive tapes. The thickness of hBN was confirmed by atomic force microscopy. The hBN/BLG/hBN/graphite heterostructure was fabricated by bubble-free transfer technique.[1] The double-dot nanostructure was patterned by electron beam (EB) lithography with 125 kV acceleration voltage, followed by reactive ion etching with $O_2$/$CHF_3$ (4/40 sccm) plasma. The etching rate is 40 nm/min. We carefully etched the heterostructure to avoid leakage to the graphite back-gate. One-dimensional edge contacting electrodes were fabricated by EB lithography and EB evaporation with Cr/Pd/Au (5/15/90 nm).[2] After the electrode fabrication, an additional hBN flake was transferred onto the device using bubble-free technique to prevent the leakage between the top-gates and the channel. The five top-gates were fabricated on the double dots by EB lithography and EB evaporation with Ti/Au (5/72 nm). There was no leakage between each gate over the range of ±2 V. The bottom graphite was used as the back-gate to control the overall Fermi level, while three top-gates LDG, RG, and CG (shown in Figure 1d in the main text) are used for locally tuning the potential. The Hall bar device was fabricated simultaneously on the same heterostructure. The channel length and width of Hall-bar are 2 μm and 1 μm, respectively (shown in Figure 1b in the main text).

2. Device characterization of the Hall bar

From the Hall bar device close to the double-dot device, we estimate the device parameters to confirm the superlattice formation and the device quality. At a base temperature of 40 mK, the Hall measurements were performed with four terminal AC lock-in techniques. Figure S1a(b) shows the longitudinal resistivity $\rho$ (transverse resistance $R_{xy}$) as a function of back gate voltage $V_{BG}$ measured at the perpendicular magnetic field $B = 0$ T (0.2 T). From the peak position and the sign change positions, we find the position of charge neutrality point (CNP) ~ 0 V and the satellite points ~±15 V for both electron and hole sides. The Hall mobility is estimated to be 110,000 cm$^2$V$^{-1}$s$^{-1}$ and 140,000 cm$^2$V$^{-1}$s$^{-1}$ for the electrons and the holes, respectively. From the width of the peak at the CNP, the residual carrier density is estimated to be ~2.0×10$^{10}$ cm$^{-2}$.[3] These indicate the high-quality in our BLG superlattice sample.

Using the distance between the CNP and the satellites, the moiré wavelength $\lambda$ can be expressed by the formula,[4]

$$\lambda = \sqrt{\frac{8e}{\sqrt{3}C_{BG}|V_{CNP} - V_{sat}|}} \cdots (1)$$

where, $e$ is the electron charge, $C_{BG}$ is the geometric capacitance for the back gate, $V_{CNP}$ is the position of the CNP, and $V_{sat}$ is the position of the satellite point. Then, we obtain $\lambda = 10.7$ nm. This value can be converted to the alignment angle $\theta$ by the equation,[5]

$$\lambda = \frac{(1+\delta)a}{\sqrt{2(1+\delta)(1-\cos\theta) + \delta^2}} \cdots (2)$$

where, $\delta = 1.8\%$ is mismatch of the lattice constant of graphene and hBN, a = 2.46 Å is the lattice constant of graphene. Then, we obtain $\theta \sim 0.85°$.

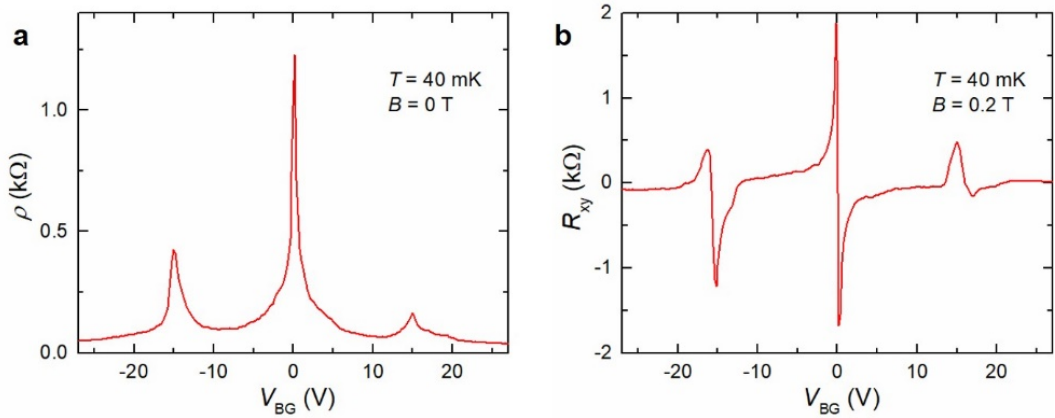

**Figure S1.** Hall bar device characteristics. (a) $\rho$ vs $V_{BG}$ measured at $T = 40$ mK, $B = 0$ T. (b) $R_{xy}$ vs $V_{BG}$ measured at $T = 40$ mK, $B = 0.2$ T.

3. Temperature dependence of the Hall bar

To estimate the energy gap from the Arrhenius analysis, we measure the $T$ dependence of the transport properties of the Hall bar device from 3 K to 15 K. Figure S2a shows $\rho$ of the Hall bar device as a function of $V_{BG}$ near the CNP. The peak broadens and its height is lowered with increasing $T$. Near the hole-side satellite as shown in Figure S2b, the satellite peak is also lowered with increasing $T$. These results indicate semiconductor-like behavior. Figure S2c,d shows the Arrhenius plot for the CNP ($\rho_{CNP}$) and the hole-side satellite ($\rho_{sat}$). For the data for high $T$ region, we fit the thermal activation transport model,

$$\frac{1}{\rho} \approx \exp\left(-\frac{\Delta}{2k_B T}\right) \cdots (3)$$

where, $\Delta$ is the activation energy, $k_B$ is the Boltzmann constant. As a result, the activation energy at the CNP is estimated to be $\Delta \sim 0.44$ meV. For the hole-side satellite peak, as a result of the same method at the CNP, the activation energy is $\Delta \sim 0.28$ meV. These results are inconsistent with the theoretical prediction.[6] However, the prediction is for the superlattice with perfect angle alignment $\theta = 0°$, and the energy gap is abruptly dropped when the angle alignment deviates from the $\theta = 0°$.

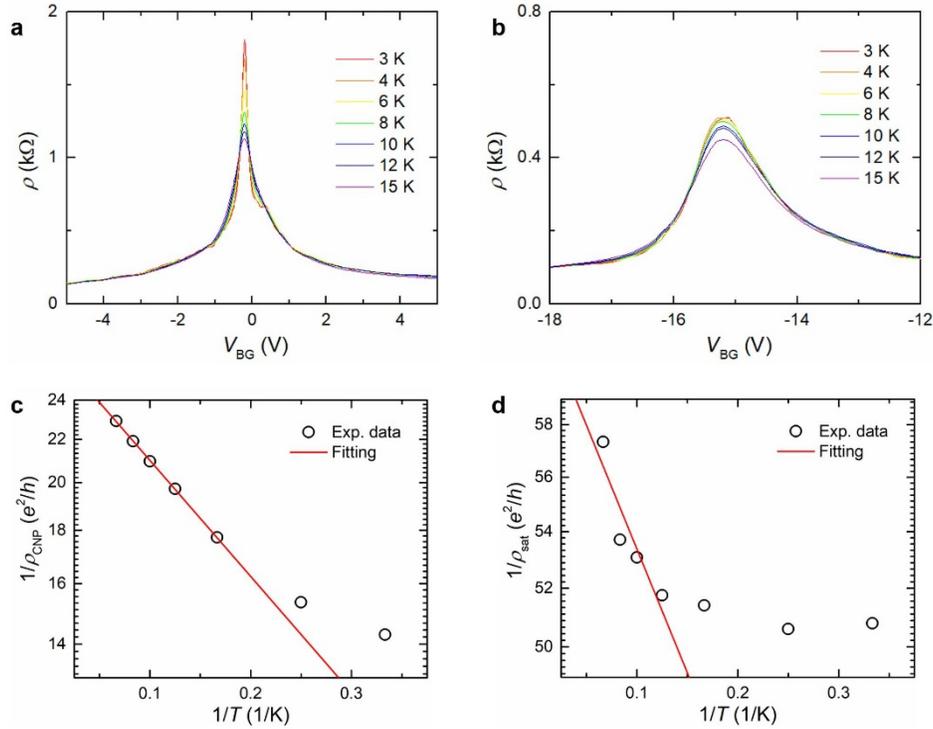

**Figure S2.** Temperature dependence of the Hall bar. (a,b) $\rho$ vs $V_{BG}$ characteristics measured for various $T =$ 40 mK near the CNP (a) and the hole-side satellite (b). (c,d) $1/\rho$ vs $1/T$ Arrhenius plots for the CNP (c) and the hole-side satellite (d) Open circle symbols are the experimental data, and the red line corresponds to the fitting result using Eq. (3).

4. Estimation of the effective electron temperature

Figure S3a shows one of the normalized Coulomb peaks for $T = 100$ mK ~ 1 K at $V_{SD} = 10$ μV. The peak width is broadened with increasing $T$. As shown in Figure S3b, we estimate the effective electron temperature $T_e$ from the peak at $T = 40$ mK by the orthodox Coulomb blockade model,[7]

$$G \approx \frac{1}{2}\cosh^{-2}\left(\frac{\alpha_{BG}|V_{BG} - V_{peak}|}{2.5 k_B T_e}\right) \cdots (4)$$

where, $\alpha_{BG}$ ~ 0.0575 is the back-gate conversion factor estimated from the Coulomb diamonds by sweeping $V_{BG}$, $V_{peak}$ is the position of the peak center. Then, we obtain $T_e$ ~ 160 mK, which almost agrees with the base temperature.

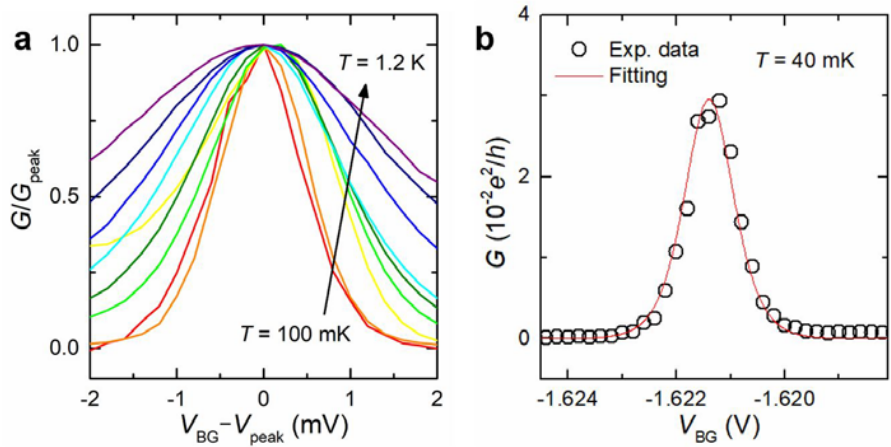

**Figure S3.** Temperature dependence of the Coulomb peak. (a) One of the Coulomb peaks for various $T$ from 100 mK to 1.2 K. Conductance $G$ is normalized with the peak value $G_{peak}$. Each peak is shifted to the center. (b) Fitting the orthodox Coulomb blockade model to the Coulomb peak at $T = 40$ mK and $V_{SD} = 10$ μV. Open circle symbols are the experimental data, and the red line corresponds to the fitting result using Eq. (4).

5. Estimation of the double-dot parameters

The double-dot parameters are derived by using the pure capacitance model.[8] Figure S4a,b shows the zoom up of the honeycomb structure in the charge-stability diagram (in Figure 3a in the main text), we extract the double-dot parameters. The results are summarized in Table S1. The left-dot and right dot have similar parameters, in good agreement in the symmetric device geometry (Figure 1c in the main text).

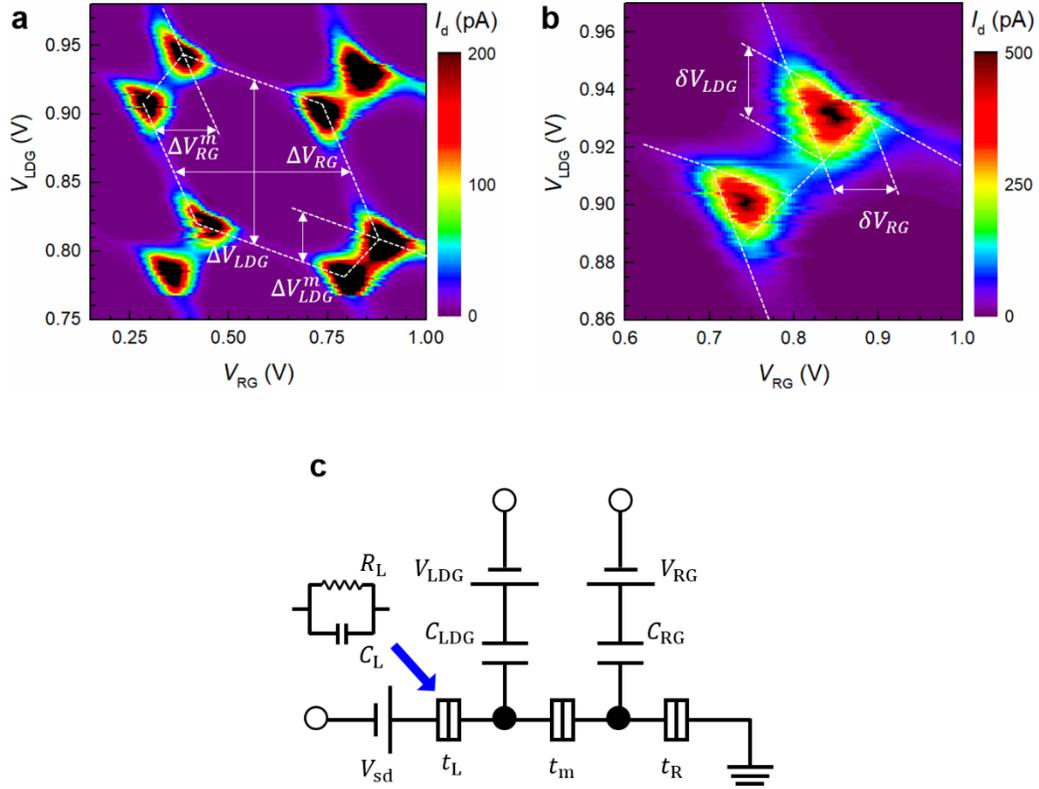

**Figure S4.** Charge-stability diagram for extracting the double-dot parameters. (a) Zoom up of the charge-stability diagram shown in Figure 3a in the main text. Scales are used for extraction of the double-dot parameters. (b) Zoom up of the bias triangles. **c**, Equivalent circuit model of the double-dot device.

**Table S1.** Double-dot parameters. Corresponding components are shown in Figure S4c.

| Components | $C_{LDG}$ | $C_{RG}$ | $C_{\Sigma L}$ | $C_{\Sigma R}$ |
|---|---|---|---|---|
| Extracted values | 1.3 aF | 0.36 aF | 110 aF | 190 aF |

6. $I_D$-$V_{BG}$ curves at the fixed perpendicular magnetic field near the charge neutrality point

In the transport properties of the double-dot device under $B$, we observe the region where the Coulomb oscillation and current are suppressed. For clarity, the cutting-line-plot is shown in Figure S5. It is obvious that the suppressed region expands with increasing $B$.

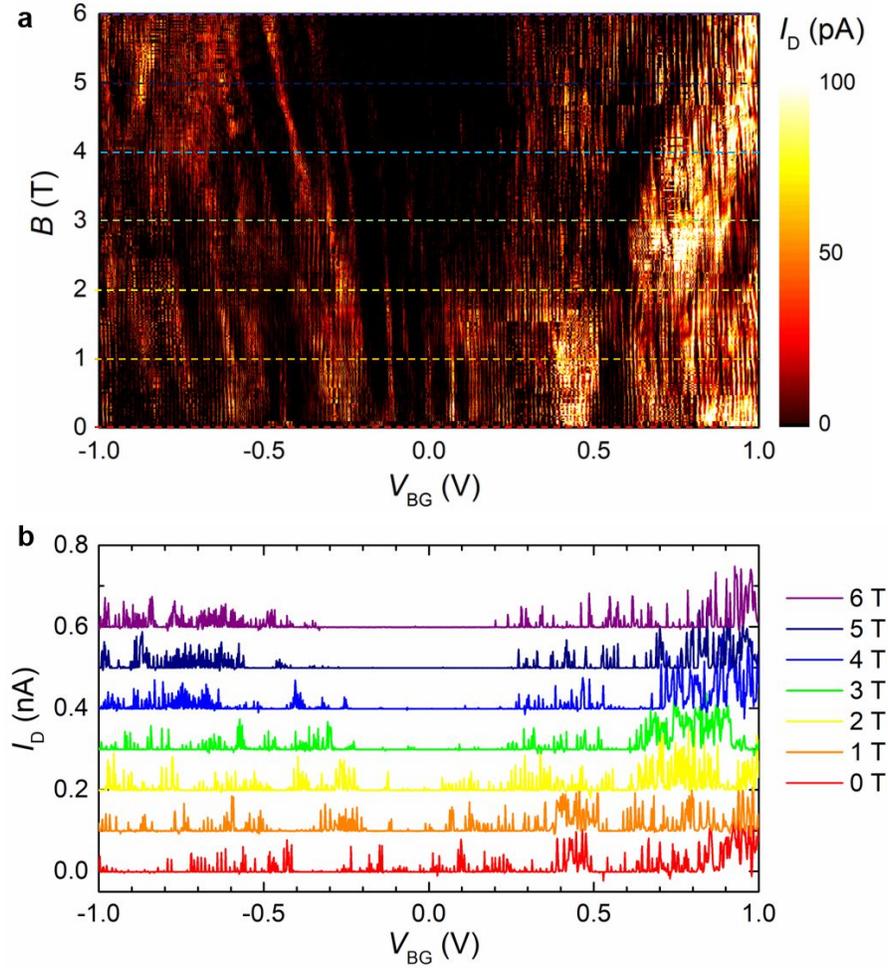

**Figure S5.** Transport property of the double-dot device in the perpendicular magnetic field. (a) $I_D$ fan diagram mapping for $V_{SD}$ = 10 μV near the CNP (the same plot as Figure 5b in the main text). (b) $I_D$-$V_{BG}$ curves for $B$ = 0-6 T. Each curve has offset of 0.1 nA for clarity. Colors of the dashed lines in (a) correspond to that of the curves in (b).

7. Transport property in a parallel magnetic field

To gain a better understanding of the current suppression region near the CNP under the perpendicular magnetic field, we measure the transport properties of the double-dot device in a parallel magnetic field $B_\parallel$. Note that this measurement was performed in a different cooling cycle due to the limitation of our equipment. As shown in Figure S6, neither the Hofstadter's butterfly spectrum for the wide $V_{BG}$ range nor the current suppression region near the CNP are not observed in contrast to the behavior under the perpendicular magnetic field. This result supports the discussion in the main text and suggests that the current suppression region could be due to the Landau level gap.

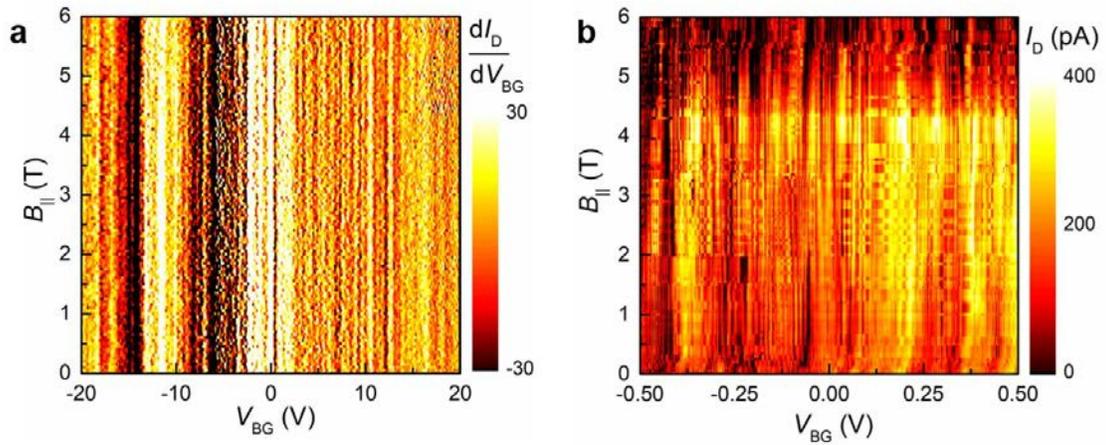

**Figure S6.** Transport property of the double-dot device in the parallel magnetic field. (a,b) $V_{BG}$-$B_\parallel$ mapping of $I_D$ for $V_{SD}$ = 10 mV for the wide $V_{BG}$ range (a), and $V_{SD}$ = 10 μV near the CNP (b).